# Intrusion Response Systems: Past, Present and Future

**Pushpinder Kaur Chouhan[1], Alfie Beard[1], and Liming Chen[2]**
[1]BT Plc, UK
[2]Department of Computer Science, Ulster University, Northern Ireland

Corresponding author: Pushpinder Kaur Chouhan (e-mail: Pushpinder.Kaur.Chouhan@bt.com).

This research is supported by BTIIC (BT Ireland Innovation Centre), funded by BT and Invest Northern Ireland

**ABSTRACT** The rapid expansion of the Internet of Things and the emergence of edge computing-based applications has led to a new wave of cyber-attacks, with intensity and complexity that has never been seen before. Historically most research has focused on Intrusion Detection Systems (IDS), however due to the volume and speed of this new generation of cyber-attacks it is no longer sufficient to solely detect attacks and leave the response to security analysts. Consequently, research into Intrusion Response Systems (IRS) is accelerating rapidly. As such, new intrusion response approaches, methods and systems have been investigated, prototyped, and deployed. This paper is intended to provide a comprehensive review of the state of the art of IRSs. Specifically, a taxonomy to characterize the lifecycle of IRSs ranging from response selection to response deployment and response implementation is presented. A 10-phase structure to organize the core technical constituents of IRSs is also presented. Following this, an extensive review and analysis of the literature on IRSs published during the past decade is provided, and further classifies them into corresponding phases based on the proposed taxonomy and phase structure. This study provides a new way of classifying IRS research, thus offering in-depth insights into the latest discoveries and findings. In addition, through critical analysis and comparison, expert views, guidance and best practices on intrusion response approaches, system development and standardization are presented, upon which future research challenges and directions are postulated.

## I. INTRODUCTION

In Internet of Things (IoT) environments with critical infrastructure (e.g. built environment integrity, utility networks, and flood management), security and reliability are paramount. Taking inappropriate actions or failing to take appropriate actions in the event of cyberattacks can have a significant cost and be extremely damaging. The challenge is to respond at the right time, with an appropriate action(s) that helps to mitigate the attack and return the IoT environment back to normal operation. To address this problem, decision-making systems for selection and initiation of appropriate actions are needed for effective and efficient utilization of IoT environments. One such system that has been in use for network intrusion for decades and is being improved with the availability of new technologies and advancement of computing power is Intrusion Response System (IRS) [1].

IRS mitigates the intrusions detected by Intrusion Detection System (IDS) [2]. A network intrusion is any unauthorized activity on a computer network. In most cases, such unwanted activity absorbs network resources intended for other uses, and nearly always threatens the security of the network and/or its data. Properly designing and deploying an intrusion detection system will help block intruders and detect intrusions that can be resolved by IRS. Response includes several stages, including preparation for incidents, detection and analysis of a security incident, containment, eradication, and full recovery, and post-incident analysis and learning.

Thus, organizations have incident response and management teams. They are required to develop and implement Incident Response Plans (IRP) in order to quickly discover attacks, before effectively containing any damage, eradicating the attacker's presence and restoring the integrity of the network and systems. The IRP should give details about the structure of the incident response team, members of the emergency response team, roles and responsibilities of the incident response team, muster points and the decision-making processes and escalation steps. Many organizations [1-4] have provided guidelines and best practices for intrusion response.

There are various survey articles [1, 2, 7-11, 80, 109, 121, 125] that reference the IRS phases and technologies, but none has presented all the phases of IRS process flow and the different techniques that can be used to address these phases. [1] presents the Phylogenetic tree for intrusion response approaches for automated IRSs: adaptive-based, expert-based and association-based. [2] provides in-depth intrusion detection systems and types of attacks along with the response selection [64, 65] process based on the different types of network attacks. [7] proposes a taxonomy for only six of the related themes of IRSs, namely time of detection, type of attack, type of attacker, degree of suspicion, attack implications and environmental constraints. [8] presents only five phases of IRSs: response cost, adjustment ability, response selection, response execution and target. [9] presents six phases of IRSs: degree of automation, activity of triggered response, ability to adjust, time of response, cooperation ability and response selection method.

[10] presents an investigation and survey of response options (passive, proactive and reactive response) based on the attack timeframe. [11] has classified the response systems into four categories - response through static decision tables, response through a dynamic decision process, intrusion tolerance through diverse replicas and intrusion response for specific classes of attacks. [80] provided response and reconfiguration of cyber-physical control systems. [120] proposed a decision-making framework for IRS considering only four metrics (i.e., attack damage, deployment cost, negative impact on QoS, and security benefit). [121] categorized IRSs into four classes (static decision-making IRSs, dynamic decision-making IRSs, intrusion tolerance through diverse replicas IRSs and IRSs meant to target specific kinds of attacks). However, none of these survey articles presents a complete cycle of response deployment along with the important features of IRSs.

The unique aspects of this article are: 1) present the methods used by existing IRSs from the past decades in a tabular summary; 2) introduce a complete life cycle of IRSs in 10-phases; 3) include the latest IRS developments in various phases, e.g., data driven IRSs; 4) establish a methodology for organizing the IRS literature according to these 10-phases; 5) offer insights and directions for future research.

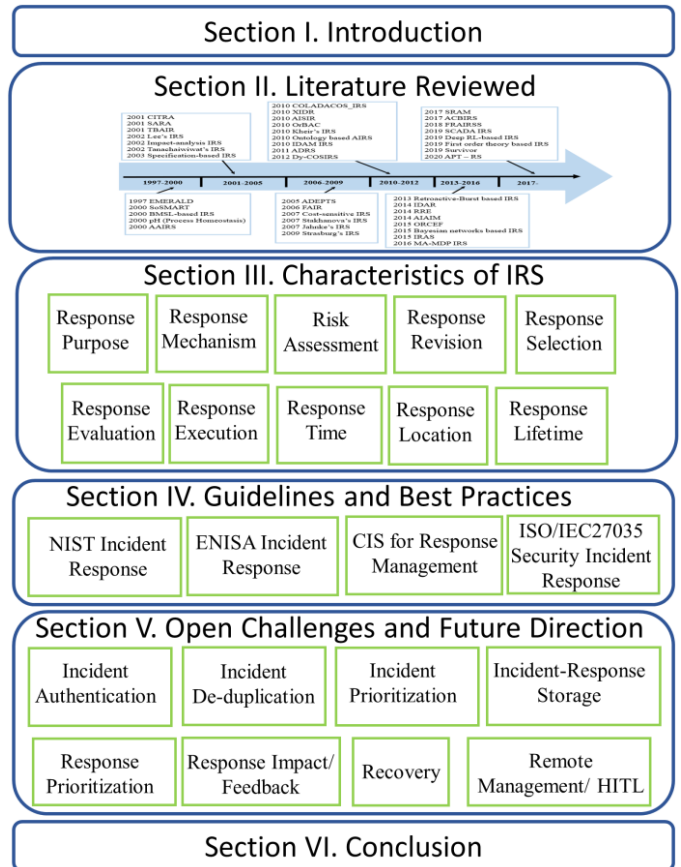

**FIGURE 1. Overview of "Intrusion Response System: A Study of Past, Present and Future"**

### A. AIM OF THIS PAPER

If your paper is intended for a conference, please contact your conference editor concerning acceptable word processor formats for your particular conference. The main goal of this paper is to survey the literature on IRS in order to provide a comprehensive reference point of the approaches used for response selection. This will help to identify the gaps that make current IRSs vulnerable and less efficient. The paper will also provide a taxonomy of IRSs based on response characteristics, such as response selection (decision making methods), response deployment (degree of automation), response implementation (time to respond), etc.

This survey paper focuses on the analysis of various methods developed from different phases of IRSs. IRS lifecycle phases have been considered as the base of this article rather than a performance or complexity comparison of different IRSs, since existing IRSs perform in diverse application scenarios using different datasets and in different real world environments. Thus, such an analysis will struggle to provide any meaningful or useful insights into the complexity/performance of different methods/approaches for each of the IRSs.

Moreover, this paper presents a comprehensive survey of research trends in the IRS lifecycle, IRS frameworks with features and open challenges for IRSs. Furthermore, it highlights the need for standardizing intrusion response processes, since it can be argued that this is a critical step towards creating best practices in security response.

This survey aims to address the following questions:

i) What are the main characteristics of an IRS?

ii) Which techniques have been established and are deployed for handling each characteristic?

iii) Are there any features or techniques that could improve the performance of IRSs, in terms of efficiency and effectiveness?

iv) What are the standards, regulations, and best practice guidelines, which could help to further improve/develop IRSs?

v) What are the open challenges for IRSs and how can they be resolved?

### B. BASIC TERMS

The definition of common/confusing terms are described below to make the reading of this paper easy; however, the following descriptions are generally applicable:

- An incident is any adverse event whereby some aspect of host/environment security could be threatened: loss of data confidentiality, disruption of data or system integrity, or denial of availability. Incidents can be sub categorized as alerts or attacks (intrusion).
- An intrusion or an attack is any unauthorized activity on a computer network, host or environment that causes interruption to a service, or reduction in the quality of a service. For example, when an intruder penetrates the security of a system.
- Alerts are notifications of events that may signify the presence of an attack or indication of future attack.
- A response is a plan for handling an incident methodically. If an incident is nefarious, steps are taken to quickly contain the attack, and learn from the incident.
- An action is any activity taken to respond to an incident.
- A taxonomy is a system with associated rules for classifying events into categories.
- Cybersecurity is the ability of network and information systems to resist action that compromises the availability, authenticity, integrity or confidentiality of digital data or the services those systems provide.
- Network and information system are an electronic communications network, or any device or group of interconnected devices which process digital data, as well as the digital data stored, processed, retrieved or transmitted.
- Incident response is a structured process to identify and deal with cybersecurity incidents.
- Incident response plan is the guide for how to react in the event of a security breach with a goal to minimize damage, reduce disaster recovery time and mitigate breach-related expenses,
- Incident response cycle is the sequence of phases that a security event goes through from the time that it is identified as a security compromise (incident) to the time that it is resolved and reported.

### C. ABBREVIATIONS

| Acronym | Full form |
|---|---|
| CERT | Computer Emergency Response Team |
| CIA | Confidentiality, Integrity, Availability |
| CIS | Center for Internet Security |
| CSIRP | Computer Security Incident Response Plan |
| CSIRT | Computer Security Incident Response Team |
| DDA | Data Driven Approach |
| ENISA | European Network and Information Security Agency |
| ICT | Information and Communication Technologies |
| IDRS | Intrusion Detection and Response System |
| IDS | Intrusion Detection System |
| IEC | International Electrotechnical Commission |
| IoT | Internet of Things |
| IPS | Intrusion Prevention System |
| IRP | Intrusion Response Plan |
| IRS | Intrusion Response System |
| ISMS | Information Security Management System |
| ISO | International Organisation for Standardisation |
| KDA | Knowledge driven approach |
| NIST | National Institute of Standards and Technology |

### D. STRUCTURE OF THIS PAPER

The literature survey aims to develop an in-depth understanding of the various phases of IRSs and the numerous techniques that have been used to accomplish the functionality of these phases. By establishing an in-depth understanding, new knowledge, observations and insights can be drawn to inform the research community, in particularly new researchers coming into this research area. To achieve this, a large body of the latest research on IRSs has been reviewed, the core characteristics of a typical IRS have been analysed and a new taxonomy for grouping the IRS phases (into 10-core phases and various techniques) has been developed. This literature review also highlights the gaps and open challenges in existing IRSs, suggesting resolutions and new directions, and existing guidelines and standards provided by international standard organizations are studied.

The paper is structured as follows: Section II presents a survey of the last decade's IRS literature [1,2,7-55]. Section III adds context to the work by introducing the characteristics of IRSs as 10-core phases. Section IV presents the IRS standards, guidelines and best practices provided by NIST

[4,66], ENISA [3,56-59], ISO [60] and CIS [61]. Open challenges and future research directions of IRSs are presented in Section V. The paper is then concluded in Section VI. An overview of the structure of this paper is presented in Fig. 1.

## II. LITERATURE REVIEWED

Intrusion response systems have been developed over the past several decades. Nevertheless, with the constant expansion of computer networks, such as cloud, edge computing and IoT, coupled with the ever-growing development of applications and technologies, attacks have also become more sophisticated. Intrusion response systems need to be constantly upgraded in order to tackle these new attacks. This section presents the latest intrusion response systems that have been studied and developed to the time this manuscript was written.

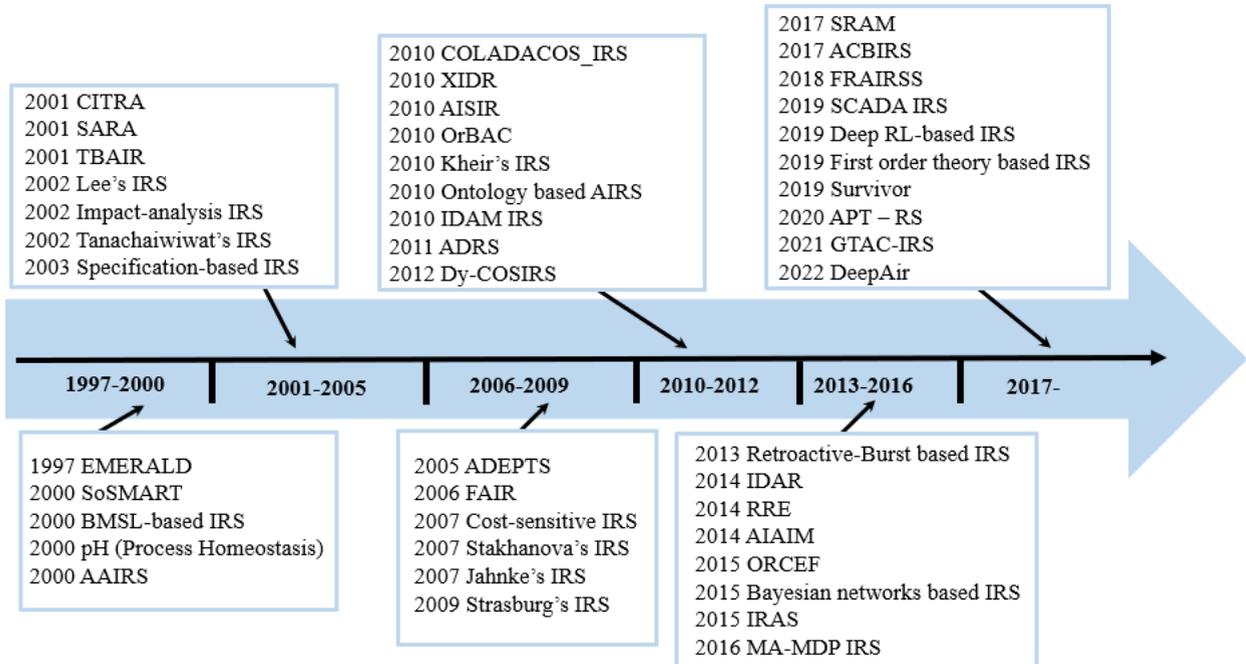

**FIGURE 2. Timeline of Intrusion Response System**

TABLE I PROBLEM ADDRESSED AND SOLUTIONS PROVIDED BY PAST INTRUSION RESPONSE SYSTEMS

| IRS | Problem Addressed | Proposed Solution |
|---|---|---|
| EMERALD [14] | How to track malicious activity through and across large networks? | Proposed an Event Monitoring Enabling Responses to Anomalous Live Disturbances (EMERALD) distributed scalable tool that deploys polymorphically at various abstract layers in a large distributed network. This uses a building-block approach to network surveillance, attack isolation, and automated response. The monitoring contributes to a streamlined event-analysis system that combines signature analysis with statistical profiling to provide localised real-time protection of the most widely used network services on the Internet. EMERALD introduces a versatile application programmers' interface that enhances its ability to integrate with heterogeneous target hosts and provides a high degree of interoperability with third-party tool suites. |
| SoSMART [15] | How to provide an automatic, adaptive response capability that provides 24/7 monitoring and responds to system and security functions across a network of computing systems? | Presented a "System or Security Managers Adaptive Response Tool" (SoSMART) that combines an agent architecture and case-based reasoning (CBR) on top of available system management or security tools. The agent architecture is used for tool integration, functional abstraction and as a medium for distributed reasoning. The CBR is used to define incident/response pairings that can recognise situations that require response and associate response actions with those situations. |
| BMSL- | How to build a | Proposed an approach to build survivable systems that combine early attack |

| based IRS [16] | survivable system? | detection with automated reaction for damage prevention and containment, as well as tracing and isolation of the attack entry point(s). This approach is based on specifying security-relevant behaviours using patterns over sequences of observable events, such as a processes system calls and their arguments and the contents of network packets. This approach can trace the origin of the attack and isolate the attacker. |
|---|---|---|
| pH [17] | How to preserve systems own integrity through automated response mechanisms? | Proposed a system called pH (for process homeostasis) which can successfully detect and stop intrusions before the target system is compromised. This system monitors every executing process on a computer at the system-call level and responds to anomalies by either delaying or aborting system calls. |
| AAIRS [18] | How to provide more robust protection to systems? | Adaptive Agent-based Intrusion Response System (AAIRS) provides response adaptation by favourably weighting the responses that have been successful in the past over the responses that have been unsuccessful. It also adapts responses based on the system's belief that intrusion detections reported are valid. Intuitively, adaptive detection and response systems provide more robust protection. |
| CITRA [19] | How to trace an intrusion back to an attacker? | A Cooperative Intrusion Traceback and Response Architecture (CITRA) is presented as a way to integrate network-based intrusion detection systems, firewalls and routers in order to trace attacks back to their true source and block the attacks close to that source. CITRA policy mechanisms integrate diverse security technologies to improve system defence. |
| SARA [20] | How to defend information systems using automated responses? | This solution proposed a Survivable Autonomic Response Architecture (SARA) with two separate "loops" of response: a local autonomic response and a global response carried out cooperatively by the hosts in a system. The primary focus of the work is to defend information systems using coordinated autonomic responses. |
| TBAIR [21] | How to identify network-based intruders? | The Tracing Based Active Intrusion Response (TBAIR) system proposed is based on Sleepy Watermark Tracing (SWT). TBAIR is able to effectively trace a detected steppingstone intrusion (to disguise its origin in real-time) and dynamically push intrusion countermeasures such as remote monitoring, blocking, containment and isolation of resources close to the source of the intrusion. TBAIR helps to apprehend the intruders on the spot and hold them accountable for their intrusions. |
| Lee's IRS [22] | How to relate intrusion detection and intrusion response cost to select appropriate responses to an intrusion? | This IRS examines the major cost factors associated with an IDS, which includes development cost, operational cost, damage cost due to successful intrusions and the cost of manual and automated responses to intrusions. These cost factors can be quantified according to a defined attack taxonomy and site-specific security policies and priorities. Present cost models to formulate the total expected cost of an IDS use cost-sensitive machine learning techniques that can produce detection models that are optimised for user-defined cost metrics. |
| Impact-analysis IRS [23] | How to evaluate the severity of an attack, so that the response causes less damage than the actual attack? | This solution presented a network model and an algorithm to evaluate the impact of response actions on the entities of a network. An evaluation algorithm compares the intrusion severity and response cost. This allows the IRS to select a response from several alternatives which fulfil the security requirements and have minimal negative effect on legitimate users. |
| Tanachaiwiwat's IRS [24] | How to collectively assess risk of multiple attacks on a network in order to automate response? | Proposes an adaptive alarm/defence framework to protect network resources by deploying desired security functions to routers, switches, firewalls and hosts. Adaptive response strategies are suggested based on the alarm confidence, attack frequency, assessed risks and estimated response costs. Dynamic security is achieved by frequent policy updates against changing attack patterns or varying network conditions. The recorded false alarms and detection rate are used in the response decision process. |
| Specification-based IRS [25] | How to reduce wasting resources while automating responses to intrusions? | This IRS proposes a specification based automated response system for handling intrusions. The system consists of an Automated Response Broker (ARB) that uses an action cost model to decide a response strategy. Even in the presence of uncertainty the system can design optimal response strategies to |

| | | mitigate an attack. |
|---|---|---|
| ADEPTS [26] | How to provide automated response to restrict the effect of intrusions on a subset of the entire set of services in distributed a system? | This solution proposes a distributed and adaptive intrusion response model using attack graphs, which evaluate the success or failure of the deployed response using a feedback mechanism. The model focuses on enforcing containment within the system, thus localising the intrusion and allowing the system to provide service, albeit degraded. |
| FAIR [27] | How to automate intrusion responses? | Flexible automated intelligent responder (FAIR) proposes an architecture to provide easily customisable response policies and adapts decisions according to changes in the environment. FAIR is based on the concept of a centralised response manager, which handles the monitoring of a number of networked client systems. Incidents that are detected on a client system trigger the initiation of appropriate response actions. |
| Stakhanova's IRS [28] | How to pre-emptively deploy responses in an intrusion response system? | Stakhanova's IRS proposes a cost-sensitive, pre-emptive, adaptive and automated model for intrusion response. The model also allows for the adaptation of responses in a changing environment by evaluating the success or failure of previously triggered responses. |
| Cost-sensitive IRS [29] | How to calculate the damage and response cost for a distributed attack in a cost-sensitive intrusion response system? | This solution proposes a cost-sensitive method to distributed intrusion response. The calculated damage and response costs reflect the cooperative intrusions based on number of attacks and number of attack sources. |
| Jahnke's IRS [30] | How to quantify the effects of a response after its deployment in an intrusion response system? | The IRS proposes using graph-based metrics for intrusion response measures. Graph-based metrics quantify relevant properties of a response measure, after its application and estimate these properties for all available response measures prior to their application. This is used as the basis for the election of an appropriate response to a given attack. |
| Strasburg's IRS [31] | How to define a consistent and adaptable measurement of cost factors on the basis of system requirements and policy for a cost-sensitive IRS? | Strasburg's IRS presents a host-based framework for cost-sensitive assessment and selection of responses. The framework is based on a set of measurements that characterise the potential costs associated with the intrusion handling process. The IRS proposes an intrusion response evaluation method with respect to the risk of potential intrusion damage, the effectiveness of the response action and the response cost on a system. |
| COLADA COS_IRS [32] | How to evaluate how much the response for an intrusion is worth? | This solution proposes a cost-sensitive model for intrusion response analysis. It considers the trade-off between cost factors such as intrusion cost, response cost and operational cost. |
| XIDR [33] | How to deploy responses at various layers in the network? | XIDR proposes a cross-layer intrusion detection and response framework, which utilises multi-source intrusion detection systems to enable cross layer intrusion detection and response deployment in a wired environment. Cross-layer automated IRSs deploy cost effective and efficient pre-emptive, as well as adaptive, responses to an ongoing intrusion. |
| AISIR [34] | How to provide artificial immunity to systems as part of an intrusion response system? | The intrusion response system proposed is based on an artificial immune system, which selects responses according to an intrusion's quantitative description and the response cost and response benefit. |
| OrBAC [12] | How to specify responses for ongoing attacks in real-time and deactivate the response once it is no longer needed? | This solution proposes a risk-aware framework for activating and deactivating policy-based responses for intrusion response systems. |
| Kheir's IRS [35] | How to bridge the gap between vulnerability graphs and exploit graphs and service dependency models of | Kheir's IRS proposes a service dependency model for cost-sensitive intrusion response systems, enabling intrusion and response impact evaluation by calculating a return-on-response-investment index. |

| | intrusion response systems? | |
|---|---|---|
| Ontology-based AIRS [36] | How to characterise response metrics for an ontology-based automated intrusion response system? | This solution presents a Semantic Web Rule Language (SWRL)-based reasoning framework to infer the most suitable response according to a set of response metrics. These metrics specify different rules for selecting a specific response, according to some contextual and input parameters and the weight associated with each of them. This framework solves the problem of adaptability and false alarms by dynamically interpreting response metrics in order to select optimal responses. |
| IDAM IRS [37] | How to improve the decision making of intrusion response systems? | IDAM IRS proposes a decision-making model based on hierarchical task network (HTN) planning. The IRS considers both response measures and response time. The model balances the response impact and response effect in a set of responses, using an online risk assessment method. |
| ADRS [38] | How to automatically and optimally balance performance objectives and potential negative consequence for anomaly detection and response systems? | This solution proposes a decision-theoretic framework to systematically analyse response cost in autonomic networks, with an objective to achieve its cost-sensitive and self-optimising operation through agent's deployment in local operating environments. The decision process is represented as a Markov chain and uses reinforcement learning to infer the optimal behaviour for deploying an adaptive and robust system. |
| Dy-COSIRS [39] | How to improve cost-sensitive response systems? | Dy-COSIRS presents a model for assessing the cost of responses based on three factors: the cost of damage caused by the intrusion, the cost of automatic response to an intrusion and the operational cost. The model dynamically adjusts response selection, based on the changing environment, during an attack by considering the effectiveness of previous actions and the feedback received. |
| Retroactive-Burst based IRS [40] | How to use historical knowledge of attacks and responses to improve an intrusion response system? | This IRS proposes an adaptive and cost-sensitive approach to response by utilising a risk assessment component to measure the effectiveness of the applied response. |
| IDAR [41,72] | How to respond to network layer attacks in mobile ad hoc networks? | Intrusion detection & adaptive response (IDAR) proposes a mechanism that employs a combination of both anomaly-based and knowledge based intrusion detection techniques and takes advantage of both techniques for protection against a variety of different attacks. IDAR deploys a flexible response scheme that depends on the measured confidence in an attack, the severity of an attack and the degradation in network performance. |
| RRE [42] | How to reduce the damage caused by an attack resulting from a delayed response? | RRE presents a response and recovery engine (RRE) that models the security battle between an attacker and itself as a multistep, sequential, hierarchical, non-zero sum, two-player stochastic game. In each step of the game, RRE leverages a new extended attack tree structure, called the attack-response tree (ART), and received IDS alerts to evaluate various security properties of the individual host systems within the network. RRE employs fuzzy logic to calculate the network-level security metric values. RRE also accounts for uncertainties in intrusion detection alert notifications. |
| AIAIM [43] | How to improve traditional passive network security systems? | This solution proposes an immune-inspired adaptive automated intrusion response system model (AIAIM). AIAIM calculates the real-time network danger using descriptions of self, non-self, memory detector, mature detector and immature detector of the network transactions. The real time network danger evaluation equations of the host or network is calculated using detectors. Then automatically adjusts intrusion response policies according to real-time danger. |
| ORCEF [44] | How to improve the response cost evaluation for intrusion response systems? | ORCEF proposes an online response cost evaluation framework based on the network elements and resource dependencies. ORCEF takes into account the user's needs in terms of quality of services (QoS) and the dependencies of critical processes in order to evaluate the positive effect and negative impact of responses with respect to attack type. |
| Bayesian | How to handle a high | This IRS proposes an Intrusion Detection and Response System which |
| networks-based IRS [45] | volume of alerts, correlate raw alerts to find attackers intensions and find the optimal responses to attacks? | processes the generated alerts in real time, correlates the alerts, calculates the risk of the correlated alerts and models the attack scenarios and their countermeasures using the concept of Bayesian decision networks. |
| IRAS [46] | How to reduce the consequences of attacks in cloud computing in real-time? | IRAS proposes an autonomic intrusion response system that uses a utility function from economics that is formulated based on big data techniques for data analysis. This is used to select the best responses to an attack. Used map reduce over the collected data to identify signatures of known attacks. |
| MA-MDP IRS [47] | How to dynamically respond to intrusions and factor in the changing nature of the system and attackers? | This solution introduces a probabilistic model-based IRS built on a multi-agent discrete-time Markov decision process (MA-MDP), which effectively captures the dynamics of both the defended system and the attacker. This model is used to automatically compose response actions to plan a multi-objective long-term response policy in order to protect the system. |
| SRAM [48] | How to improve intrusion risk assessment to enhance Intrusion Response Systems? | This IRS proposes a State-Aware Risk Assessment model based on D-S (Dempster-Shafer) evidence theory. The IRS take inputs from Intrusion Detection Systems and system state information to improve the pertinence of evaluation. |
| ACBIRS [49] | How to improve Intrusion Response Systems? | An adaptive and cost-based intrusion response system, which selects responses on cost-based response merit is proposed. The cost-based method considers features such as the type of the attack, severity of the attack, value of the targeted host/hosts services and their data, to prioritise alerts. The system also includes response feedback supervision that allows responses to be adaptive in changing environments. |
| FRAIRSS [50] | How to deal with precise measurement and uncertainty in the judgment of each criterion and the response prioritisation for the specific category of attacks? | A Fuzzy Rule-Based Automatic Intrusion Response Selection System based on Fuzzy AHP (Analytic Hierarchy Process) is proposed as a solution to deal with precise measurement and uncertainty in the judgment of each criterion. Furthermore, Fuzzy TOPSIS (Technique for Order of Preference by Similarity to Ideal Solution) is used for response prioritisation in multi-criteria decision making. |
| SCADA IRS [51] | How to protect SCADA from destructive attacks from an actuator? | This IRS deploys hardware that performs continuous operational profiling of the actuators. The hardware operates in two modes: a passive mode (normal) means simply monitoring and an active mode (malicious) means slowly stopping the actuator and sending an alert to SCADA. |
| Deep RL-based IRS [52] | How to protect large-scale systems against large-scale attacks? | This approach is based on deep reinforcement learning. A deep Q-learning algorithm is implemented to store the parameters of the underlying neural network and a single forward pass is used to find the best action to take in a particular state. |
| Fist order theory-based IRS [53] | How to incorporate beliefs and behaviours of the attackers in response selection? | This IRS is based on a theory of mind stochastic game-theoretic approach to predict an attacker's actions with nested beliefs. A multi-step attack scenario is modelled by utilising a Bayesian attack graph. |
| Survivor [54] | How to limit the damage caused by security incidents and repair any damage? | Survivor proposes an orchestration approach for fine-grained recovery and per-service responses (e.g., privilege removal). It puts the system into a degraded state (availability of core functions only), thus preventing attackers from re-infecting the system until any necessary patches are deployed. |
| APT-RS [55] | How to respond to multi-step attack sequences and restrict follow up opportunities for an attacker? | This IRS proposes a step-wise response mechanism that first analyses prior steps of the attack sequence and their impacts and then extracts an attacker's characteristics. Finally, the IRS incorporates their influence on the next adversarial options. |
| GTAC-IRS [94] | How to guarantee the scalability and security of IoT services on extended devices in the | This IRS employs a game-theoretic-based response selection matrix, aiming to reduce training time and facilitating the convergence of the response scheme. GTAC-IRS constructs a well-designed state representation of the observed environment, a low-dimension response matrix and a simplified response |

|  | W-SDN based network? | selection policy to lower the complexity of the algorithms. |
| --- | --- | --- |
| DeepAir [122] | How to improve action selection for attack scenarios in Software-Defined Networks? | This is an adaptive intrusion response solution using on deep reinforcement learning to effectively defend against cyber-attacks in SDN. DeepAir is based on a Markov decision process (MDP) approach and formulate the related optimization problem. |
| RL training framework [123] | How to take into consideration dynamic nature of network to automatically respond? | RL training framework proposed a Model-free Reinforcement Learning approach. The need to define subjective values for significant evaluation variables such as response cost and response benefit, has been removed though the use of direct experience-based selection method to evaluation of response impact. This allows the evaluation of response impact based on differing circumstances based on differing circumstances. |

To determine the latest and most impactful literature, google search, IEEE, Springer and the Elsevier article website were searched, focusing on the terms intrusion response system(s), response system(s), intrusion response(s), response selection, response management, incident management, incident handling, security management, adaptive response(s), response tool(s), damage containment, automated response, response decision-making, dynamic decision-making, and response architecture. There were numerous articles collected from these searches. The most relevant and significant articles in terms of IRSs were selected, e.g., articles that described specific IRSs, provided a survey of IRSs or guidelines for designing IRSs. Figure 2 presents the IRSs that were selected from the researched articles. These IRSs were developed to solve different problems and provide differentiated solutions as mentioned in Table 1. Most of the IRSs focus on automated response deployment with the emphasis on improving the response selection procedure and reducing the damage caused by an attack. Almost all IRSs were tackling Network based intrusion detection, except EMERALD [14], SoSMART [15], pH [17] and TBAIR [21]. These IRSs monitor the systems activities to find anomalies which are considered intrusions and are then responded to. The majority of the IRSs only tackle the intrusion but BMSL-based IRS [16], CITRA [19] and TBAIR [21] also trace the intrusion to find the origin. An intrusion detection and response scheme for CP-ABE-encrypted IoT networks [83] and a Game Theoretic Actor Critic Based intrusion response scheme for Wireless SDN-Based IoT Networks [94] have been developed in the last few years.

## III. CHARACTERISTICS OF INTRUSION RESPONSE SYSTEMS

In this section, the basic IRS characteristics are analysed in detail. These characteristics, as depicted in Figure 3, represent the core functions of the IRS framework/architecture that help to mitigate incidents, whether through introducing a corrective response or enabling enhanced sequence of responses.

Figure 4 presents the taxonomy of all phases of an IRS along with the key characteristics, it built upon and extended existing structures from the literature for summarizing IRSs to provide a new way of organizing the literature. In addition, new methodologies, such as, knowledge-driven and data-driven approaches for the response mechanism phase have been reviewed and included and new criteria for classification and analysis, offering new insights and knowledge, have been introduced. For example, IRSs have been categorized based on response time into three types - proactive, pre-emptive and reactive. Core characteristics of an IRS are described in following paragraphs.

### A. Response purpose

Each IRS tackles different security risks differently and thus uses different technologies to adhere to "The Big Three" issues of security: confidentiality (C), integrity (I), and availability (A), also known as the CIA triad. In this article, all the objects in the context of security are considered to be data, thus the CIA assets are mentioned as data. Data should be secured at each of its stages: data-in-flight, data-at-rest and data-in-use. Systems should be safeguarded for use by authorized users only. Services should be procured for accessibility by certified customers. CIA ensures that data, systems, services and resources are protected from unauthorized viewing and access. To define CIA only the data term is considered, which refers to any object that is valuable to the user. *Confidentiality* assures that no unauthorized individuals can read the data that needs to be protected. *Integrity* assures that no unauthorized individuals can delete or modify the data and also ensures that when an authorized person makes a change that should not have been made the damage can be reversed. *Availability* assures that authorized users should always have timely and uninterrupted access to its data.

### B. Response mechanism

Based on the response mechanism IRSs are categorized into two groups: *Active* IRSs and *Passive* IRSs. Active IRSs automatically respond to intrusions without any intervention required by security experts, whereas passive IRSs inform security experts of the response and do not perform any protective or corrective functions on their own.

The passive response exposes the information assets to attackers while the attacks are being investigated. *Notification passive* IRSs generate notifications on intrusions by triggering alarms and generating alert reports with attack information. *Manual passive* IRSs launch pre-configured responses when a problem arises. Passive responses notify and alert other parties to the existence of an intrusion and it's up to these parties to take further actions. An active IRS immediately produces an automated action to try and mitigate the incident without human involvement. Passive IRSs are insufficient and unable to respond to high-speed attacks because of their non-active nature. Thus, they are not useful for large number of attacks throughout a network or on sensors as seen in the enterprise security and Internet of Things domains respectively. Based on available incident information, an active IRS can be built on *knowledge driven* or *data driven* approaches (Figure 3). Active knowledge/data-driven approaches automatically deploy the response as the intrusion is detected and response is matched from the knowledge-base, unlike in passive responses where the response is only notified and not deployed.

Knowledge Driven Approaches (KDA) are an adaptive approach which provide collective wisdom from historical experiences by matching current incidents to relevant prior incidents. KDAs are used as a guideline for action selection. KDAs enable the thought process of a decision to be specified and they support the reasons behind a decision. This means that for an incident the overall outcome can be evaluated thoroughly and implemented with ease. KDAs can be categorized into three groups: *adaptive-based, expert-based* and *association-based*. Knowledge driven adaptive based IRSs use feedback loops to evaluate previous responses and revise for future intrusion responses. Knowledge driven expert-based IRSs choose the response action according to one or more metrics. Knowledge driven association-based IRSs use a simple decision table approach, wherein a specific response is associated with a specific attack.

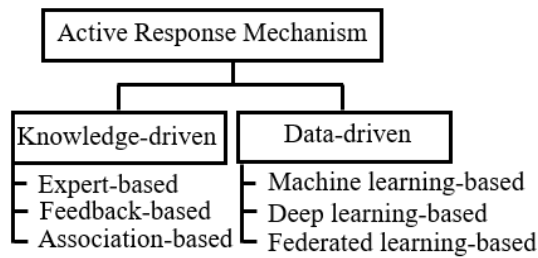

FIGURE 3. Active Response Mechanism of Intrusion Response System

Data Driven Approaches (DDA) are an analytical process that turn data into useful information that can be used for making decisions. DDAs provide inference or predictions for the target incident based on provided (incident-specific) data. DDAs are built on mathematical models that are trained on sample data to make predictions or decisions automatically without being explicitly programmed to do so. As the backbone of a DDA is data, evolving techniques for data analysis can be used for DDAs. DDAs can be categorized into three groups: *machine learning-based, deep learning-based* and *federated learning-based*. The goal of these learning based DDAs is to use existing (or develop new) computer programs that can access data and use it learn for themselves.

With easier access to high computing power and storage facilities, data can be gathered on various situations and used for analysis and improving security [85]. This data driven approach is being used by the latest developments in IRSs. [38, 52, 82, 86, 87, 122, 123, 124] used reinforcement learning techniques for response selection. Machine learning and deep learning are used in [47, 52, 88, 89] to detect attacks and respond to the detected attacks. Some authors have used other techniques, such as game theory techniques [90-95], graphical security models [96-103], network quarantine channels [104, 105], mobile agents [106-109], analytical hierarchy process [110, 111], genetic algorithm approaches [112-115] and fuzzy logic [116-119] for IRSs.

*C. Risk assessment*

Risk assessment is the process of identifying and characterizing risk. In other words, risk assessment helps the IRS determine the probability that a detected anomaly is a true problem and could actually compromise its target. Response risk assessment helps response systems to be more intelligent in terms of preventing the problem from worsening and in returning the system to a healthy state. Response risk assessment measures the risk index for each response based on response cost and attack cost via offline or online approaches.

Two types of risk assessment based on attack cost are static (offline) and dynamic (online). IRSs that deploy *static* risk assessment assign a static value to evaluate all the resources in advance. Offline risk assessment has been reviewed in ISMS and described in many existing standards, such as NIST [4] and ISO 27001[60]. IRSs that deploy *dynamic* risk assessment evaluate a risk index related to the host or network in a real time. Online risk assessment minimizes the performance cost incurred by applying a subset of all the available sets of responses when that may be enough to neutralize the attack. Online risks assessment approaches are categorized intro three main types: attack graph-based approaches, service-dependency graph-based approaches and non-graph-based approaches. An *attack graph* is used to identify attacks and their flow paths to all critical resources in the network based on service vulnerability. In the *service dependency graph*, the CIA of services are defined for each service, thus responses are mapped to specific resources instead of being statically assigned to elementary attack steps. A *non-graph-based* approach utilizes neither a graph nor CIA model for risk assessment, instead risk analysis is performed on the basis of the information provided in the alert risk assessment component.

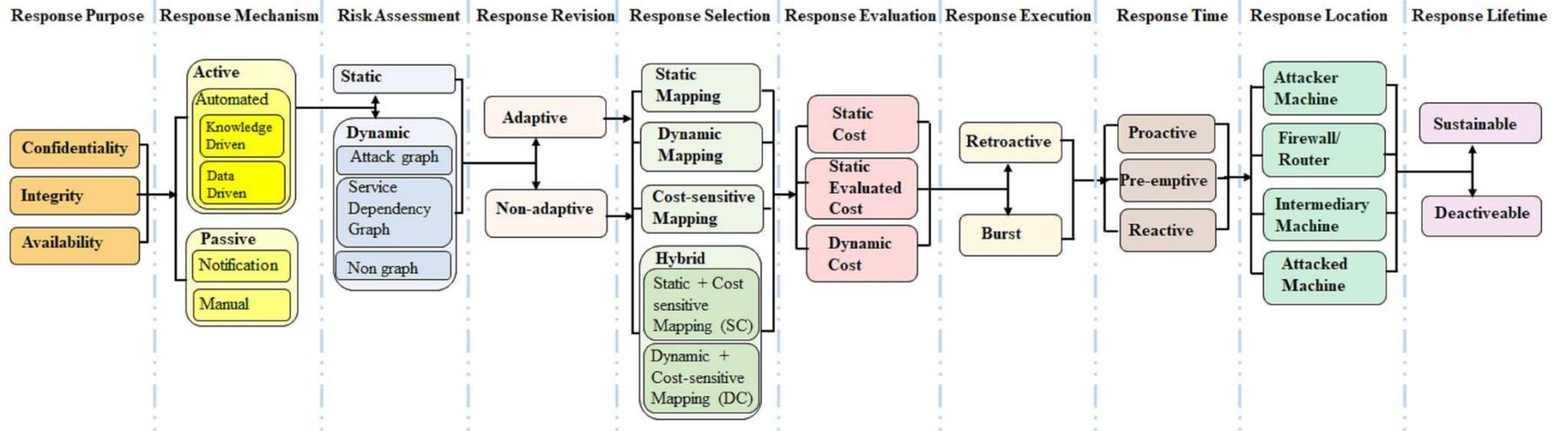

**FIGURE 4.** Taxonomy with respect to characteristics of Intrusion Response System

### D. Response revision

Response revision indicates the ability of IRSs to re-adjust the strength of the response according to the nature of attacks [82-85]. Based on their ability to revise responses, IRSs can be categorized as adaptive or non-adaptive. *Non-adaptive* IRSs always employ the same response during the response selection process for the same type of attack without considering the statistical features of the attack. *Adaptive* IRSs can dynamically adjust the response selection process according to the statistical features and response history of the attacks. [11] presented the response goodness approach to convert a non-adaptive model into an adaptive model.

### E. Response selection

Selecting an optimal response for an attack at the right time is a very crucial task. IRSs deploy different response selection methods to map appropriate responses to an attack. Based on these mapping techniques [76 - 79] IRSs can be categorized in four types: static mapping, dynamic mapping, cost-sensitive mapping and hybrid.

*Static mapping* IRSs map an attack to a predefined response. The development of static mapping IRSs is easy due to their pre-specified fixed nature but beneficial for attackers as the responses are predictable and they can learn to bypass these responses. Furthermore, this approach is inadequate for distributed large networks. *Dynamic mapping* IRSs map an attack to a predefined set of responses depending on multiple factors, such as the targeted system, system state, attack metrics (frequency, severity, confidence, etc.) and network policy. Dynamic mapping provides flexibility to an IRS because attack-response maps can be adjusted according to the attack metrics. The optimal response is dynamically chosen from a set of responses according to the statistical features of the attack. Dynamic mapping IRSs do not learn anything from attack to attack, therefore the intelligence level remains the same until the next upgrade.

*Cost-sensitive mapping* IRSs are based on a cost-sensitive metric that attempts to balance intrusion damage and response cost in the response selection process. The response is activated when the intrusion damage is greater than the response cost. The selection of the response is not based on its ability to respond to attacks but based on its effects on the target machine. Some cost sensitive approaches have been proposed that use an offline risk assessment component, which is calculated by evaluating all the resources in advance. The value of each resource is static. In contrast, an online risk assessment component can help us to accurately measure intrusion damage. The major challenge with the cost-sensitive model is the online risk assessment and the need to update the cost factor (risk index) over time. *Hybrid* IRSs combine cost-sensitive mappings with either static mappings (SC) or dynamic mappings (DC) for response selection.

### F. Response evaluation

Dynamic risk assessment with a dynamic response model offers the best response based on the current situation of the network, and so the positive effects and negative impacts of the responses must be evaluated online at the time of the attack. Evaluating the cost of the response online can be based on resource interdependencies, the number of online users, the users privilege level, etc. There are three types of response cost model: static cost model, static evaluated cost model and dynamic evaluated cost model.

The *static response cost* is obtained by assigning a static value based on expert opinion. So, in this approach, a static value is considered for each response ($RC_s$ = CONSTANT). The *static evaluated cost* is obtained by an evaluation mechanism where a statistically evaluated cost is associated with each response ($RC_{se} = f(x)$). A common method to evaluate the positive and negative effects of responses, based on the consequences for the confidentiality, integrity and availability of a system, and based on various performance metrics is to consider the consequences for other resources, in terms of availability and performance. A greater RC, indicates that the response is better: $RC_{se}$ = Positive$_{effect}$ /Negative$_{impact}$. The *dynamic evaluated cost* is based on the network situation ($RC_{de}$). The response cost is evaluated online based on the dependencies between resources and online users. Evaluating the response cost with respect to the resource dependencies, the number of online users and the user privilege level leads to an accurate cost-sensitive IRS.

### G. Response execution

When an IRS selects the response for an attack, those responses should be executed. Methods used for response execution can be categorized in two types: burst and retroactive (Figure 5).

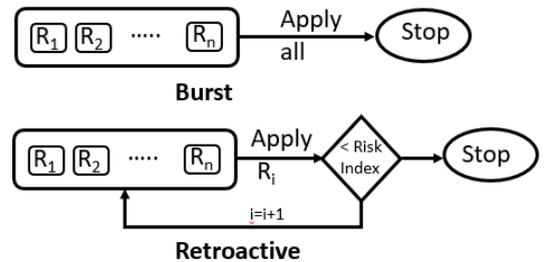

**FIGURE 5**. Response execution method for IRS

*Burst* approach apply all the responses without any appropriate measurement of the risk. Consequently, the response cost may exceed the attack cost. *Retroactive* approaches implement a feedback mechanism that measures the response effect on the basis of the most recently applied response result. The most optimal response in a set of historical responses is applied first.

### H. Response time

Different responses are necessary prior to an attack, during an attack, and after an attack. Thus, IRSs can be categorized in three types based on the response time: proactive, pre-emptive and reactive. In the *reactive* approach, all responses are delayed until the intrusion is detected. The system remains in the unhealthy state before the detection of the malicious activity and until the reactive response is applied, which gives benefit of time to attackers. Thus, the system is exposed to greater risk of damage and sometimes returning the system to a healthy state is difficult. The majority of IRSs use this approach because of the ease of response deployment, although this type of IRS is not useful for high security (real-time responses should be deployed to block the attack and mitigate its negative effects). In the *pre-emptive* approach, responses are triggered before the attack completes. Some existing solutions are only able to detect intrusions when they occur, either partially or fully, thus researchers have proposed prediction functions and alert filtering/correlation approaches with a view to predict the goals of attackers. In the *proactive* approach, responses attempt to control and prevent a malicious activity before it happens and play a major role in defending hosts and the network. Pre-emptive and proactive approaches are beneficial for multi-step attacks. Risk assessment can influence the design and planning strategies for the launch of the response, e.g., in deciding the response time (proactive, pre-emptive or reactive).

### I. Response location

The numerous locations and the variety of responses at each location will constitute a more complex framework for defending a system from attack, since its behavior will be less predictable. Broadly four locations are possible where a response can be deployed: attacker machine, firewall/router, intermediary machine and attacked machine. The *attacker machine* is the point from where the attack is generated. Deploying a response on an attacker's machine is efficient performance-wise, but not easy. *Firewalls and routers* are the firewall points. *Intermediary machines* are machines that the attacker exploits (through vulnerabilities) to compromise the target host. *Attacked machines* are the end points which the attacker targets.

### J. Response lifetime

Response lifetime is the duration for which a response is being deployed as an attack countermeasure. Based on the duration of a response deployment, the response lifetime can be categorized as sustainable and one-shot. *One-shot* response have an effective lifetime that can be defined or marked as single deployment. When a response in this category is launched, it is automatically deactivated. A *sustainable* response remains active to deal with future threats after a response in this category has been applied. Most responses are temporary actions which have an intrinsic cost or induce side effects on the monitored system, or both. Even, the deactivation of policy-based responses is not a trivial task. Thus, responses should have a specific lifetime.

Based on the taxonomy and characteristics described above, the existing IRSs presented in Section II have been categorized, as displayed in Table 2. As each IRS implements different combinations of these methods, each papers' research has been characterized and categorized through the mapping of their used techniques for each phase. 49% of the reviewed IRSs response purpose is for all 3 aspects of security, i.e., CIA. Almost all IRSs use a knowledge-driven response mechanism with the exceptions being IDAR and the Deep-RL based IRS, which use data mining and Deep Q-learning respectively. Only 4.65% of the reviewed IRSs use a data-driven response mechanism, which is a recent trend, seen in the last few years. In future more IRSs will be developed using DDAs for the response mechanism due to the availability of high computing power and storage facilities, which greatly aid machine learning techniques. Adaptive response revision is performed by 46% of the reviewed IRSs and 85% of these were IRSs designed in the last few years. Most of the IRSs use a cost sensitive mapping technique for response selection and very few IRSs have proactive and pre-emptive responses. Most IRSs deploy responses directly on an attacked machine. In addition, several IRSs have feedback loops and update the response by considering the success of the deployed response. Only 14% of the reviewed IRSs have a one-shot response lifetime. However, risk assessment and response evaluation techniques by IRSs are distributed sporadically amongst all the IRSs.

## IV. INTRUSION RESPONSE: GUIDELINES AND BEST PRACTICE

Many organizations such as the National Institute of Standards and Technology (NIST), European Network and Information Security Agency (ENISA), Computer Internet Security (CIS) and International Organization for Standardization (ISO) have provided numerous documents [3, 4,56-59, 60, 61, 66-70] on guidelines, best practice and standards for handling common network security incidents in great detail. Each organization has incident response policies, which consist of procedures that explain precisely how to respond to the most common security threat vectors and their associated incidents. In this section some of the security response documents released by these organizations are discussed.

TABLE II THE CLASSIFICATION OF IRSS BASED ON THE PROPOSED TAXONOMY

| IRS | Response Purpose | Response Mechanism | Risk Assessment | Response Revision | Response Selection | Response Evaluation | Response Execution | Response Time | Response Location | Response Lifetime |
|---|---|---|---|---|---|---|---|---|---|---|
| EMERALD [14] | Availability | Knowledge driven | Static | Non-adaptive | Dynamic mapping | Static cost | Burst | Reactive | Attacker machine | Sustainable |
| SoSMART [15] | Integrity, Availability | Knowledge driven | Service dependency | Non-adaptive | Static mapping | Static cost | Burst | Reactive | Attacker Attacked | Sustainable |
| BMSL-based IRS [16] | Integrity, Availability | Knowledge driven | Service dependency | Non-adaptive | Static mapping | Static cost | Burst | Pre-emptive Reactive | Attacker Attacked | Sustainable |
| pH [17] | Integrity | Knowledge driven | Service dependency | Non-adaptive | Static mapping | Static cost | Burst | Proactive Pre-emptive | Attacked machine | Sustainable |
| AAIRS [18] | Integrity, Availability | Knowledge driven | Non graph | Adaptive | Dynamic mapping | Static evaluated cost | Burst | Reactive | Attacked machine | Sustainable |
| CITRA [19] | CIA | Knowledge driven | Static | Non-adaptive | Dynamic mapping | Static cost | Burst | Reactive | Attacker Attacked | Sustainable |
| SARA [20] | Availability | Knowledge driven | Static | Non-adaptive | Dynamic mapping | Static cost | Burst | Reactive | Attacked machine | Sustainable |
| TBAIR [21] | CIA | Knowledge driven | Static | Non-adaptive | Dynamic mapping | Static cost | Burst | Reactive | Attacker Intermediary | Sustainable |
| Lee's IRS [22] | CIA | Knowledge driven | Static | Non-adaptive | Cost-sensitive mapping | Static cost | Burst | Reactive | Attacker Attacked | Sustainable |
| Impact-analysis IRS [23] | Availability | Knowledge driven | Static | Non-adaptive | Cost-sensitive mapping | Dynamic cost | Burst | Reactive | Intermediary/attached machine | Sustainable |
| Tanachaiwiwat's IRS [24] | Availability | Knowledge driven | Static | Non-adaptive | Cost-sensitive mapping | Static cost | Burst | Reactive | Intermediary/Attacked | Sustainable |
| Specification-based IRS [25] | CIA | Knowledge driven | Service dependency | Non-adaptive | Cost-sensitive mapping | Dynamic cost | Burst | Reactive | Attacked machine | Sustainable |
| ADEPTS [26] | Availability | Knowledge driven | Attack graph | Adaptive | Cost-sensitive mapping | Static cost | Burst | Proactive | Attacked machine | Sustainable |
| FAIR [27] | CIA | Knowledge driven | Non graph | Non-adaptive | Cost-sensitive mapping | Static evaluated cost | Burst | Reactive | Attacked machine | Sustainable |
| Stakhanova's IRS [28] | CIA | Knowledge driven | Non graph | Adaptive | Cost-sensitive | Static evaluated | Burst | Pre-emptive | Attacked machine | Sustainable |
| | | | | | mapping | cost | | | | |
| Cost-sensitive IRS [29] | CIA | Knowledge driven | Non graph | Adaptive | Cost-sensitive mapping | Static cost | Burst | Reactive | Attacked machine | Sustainable |
| Jahnke's IRS [30] | CIA | Knowledge driven | Attack graph | Non-adaptive | Cost-sensitive mapping | Dynamic cost | Burst | Reactive | Attacked machine | Sustainable |
| Strasburg's IRS [31] | CIA | Knowledge driven | Static | Adaptive | Cost-sensitive mapping | Static evaluated cost | Burst | Reactive | Attacked machine | Sustainable |
| COLADACOS_IRS [32] | Availability | Knowledge driven | Static | Adaptive | Cost-sensitive mapping | Static cost | Burst | Reactive | Attacked machine | Sustainable |
| XIDR [33] | Availability | Knowledge driven | Service dependency | Adaptive | Dynamic mapping | Static evaluated cost | Retroactive | Pre-emptive | Intermediary / Attacked machine | Sustainable |
| AISIR [34] | Availability | Knowledge driven | Static | Non-adaptive | Cost-sensitive mapping | Static cost | Burst | Reactive | Attacked machine | Sustainable |
| OrBAC [12] | CIA | Knowledge driven | Service dependency | Adaptive | Cost-sensitive mapping | Static evaluated cost | Burst | Proactive | Attacked machine | One-shot |
| Kheir's IRS [35] | Availability | Knowledge driven | Service dependency | Non-adaptive | Cost-sensitive mapping | Dynamic cost | Burst | Proactive | Attacked machine | Sustainable |
| Ontology-based AIRS [36] | Availability | Knowledge driven | Service dependency | Adaptive | Cost-sensitive mapping | Static evaluated cost | Burst | Reactive | Attacked machine | Sustainable |
| IDAM IRS [37] | CIA | Knowledge driven | Static | Non-adaptive | Cost-sensitive mapping | Static cost | Burst | Reactive | Attacked machine | One-shot |
| ADRS [38] | CIA | Knowledge driven | Non graph | Adaptive | Dynamic mapping | Dynamic cost | Retroactive | Proactive Pre-emptive | Attacked machine | Sustainable |
| Dy-COSIRS [39] | CIA | Knowledge driven | Non graph | Adaptive | Cost-sensitive mapping | Dynamic cost | Retroactive | Reactive | Attacked machine | Sustainable |
| Retroactive-Burst based IRS [40] | N/A | Knowledge driven | Service dependency | Adaptive | Cost-sensitive mapping | Dynamic cost | Retroactive | Reactive | Attacked machine | Sustainable |
| IDAR [41,73,76] | CIA | Data driven | Non graph | Adaptive | Cost-sensitive | Static evaluated | Burst | Reactive | Attacked machine | Sustainable |

| | | | | | mapping | cost | | | | |
|---|---|---|---|---|---|---|---|---|---|---|
| RRE [42] | IA | Knowledge driven | Attack graph | Non-adaptive | Dynamic mapping | Static evaluated cost | Retroactive | Reactive | Attacked machine | Sustainable |
| AIAIM [43] | Availability | Knowledge driven | Non graph | Non-adaptive | Dynamic mapping | Static cost | Retroactive | Reactive | Attacked machine | Sustainable |
| ORCEF [44] | CIA | Knowledge driven | Service dependency | Adaptive | Cost-sensitive mapping | Static evaluated cost | Retroactive | Reactive | Attacked machine | Sustainable |
| Bayesian networks-based IRS [45] | CIA | Knowledge driven | Attack graph | Non-adaptive | Dynamic mapping | Static evaluated cost | Retroactive | Pre-emptive, Reactive | Intermediary / Attacked machine | One-shot |
| IRAS [46] | CIA | Knowledge driven | Service dependency | Non-adaptive | Dynamic mapping | Static evaluated cost | Retroactive | Reactive | Attacked machine | Sustainable |
| MA-MDP IRS [47] | CIA | Knowledge driven | Attack graph | Non-adaptive | Hybrid DC | Static evaluated cost | Burst | Reactive | Attacked machine | Sustainable |
| SRAM [48] | CIA | Knowledge driven | Service dependency | Adaptive | Dynamic mapping | Static evaluated cost | Retroactive | Pre-emptive | Intermediary / attacked machine | Sustainable |
| ACBIRS [49] | IA | Knowledge driven | Service dependency | Adaptive | Cost-sensitive | Dynamic cost | Retroactive | Reactive | Attacked machine | One-shot |
| FRAIRSS [50] | CIA | Knowledge driven | Non graph | Non-adaptive | Dynamic mapping | Dynamic cost | Retroactive | Reactive | Attacked machine | Sustainable |
| SCADA IRS [51] | Integrity | Knowledge driven | Non graph | Non-adaptive | Dynamic mapping | Static evaluated cost | Burst | Pre-emptive | Intermediary machine | Sustainable |
| Deep RL-based IRS [52] | CIA | Data driven | Non graph | Adaptive | Hybrid (DC) | Dynamic Cost | Retroactive | Reactive | Attacked machine | One-shot |
| Fist order theory-based IRS [53] | Integrity, Availability | Knowledge driven | Attack graph | Adaptive | Dynamic mapping | Dynamic Cost | Retroactive | Pre-emptive | All | Sustainable |
| Survivor [54] | Integrity, Availability | Knowledge driven | Service dependency | Adaptive | Cost-sensitive | Dynamic cost | Burst | Pre-emptive | Attacked machine | Sustainable |
| APT-RS [55] | Integrity, Availability | Knowledge driven | Attack graph | Adaptive | Dynamic | Dynamic cost | Retroactive | Pre-emptive | Attacked machine | One-shot |

## A. NATIONAL INSTITUTE OF STANDARDS AND TECHNOLOGY

The National Institute of Standards and Technology (NIST) is a physical sciences laboratory and non-regulatory agency of the United States Department of Commerce, which sets standards and recommendations for many technology areas, including cybersecurity.

The NIST Computer Security Incident Handling Guide [4] provides in-depth guidelines on how to build an incident response capability within an organization. It covers several different models for incident response teams, advice on how to select the best model and information on best practices for managing an incident response team. The guide also highlights the post-incident activities that should be conducted, including information on the lessons learnt, evidence retention and using collected incident data. Furthermore, NIST has grouped the incident response process into four phases as shown in Figure 6.

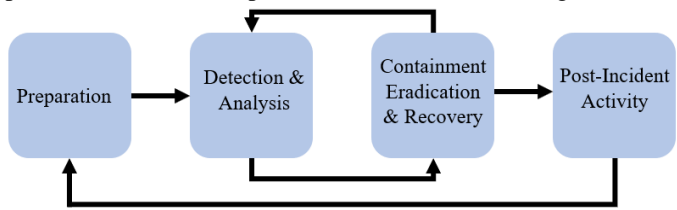

FIGURE 6. *NIST Incident Response Life Cycle*

The NIST incident response process contains four steps: (1) Preparation, (2) Detection and Analysis, (3) Containment, Eradication and Recovery, and (4) Post-Incident Activity.

*Preparation* is key to rapid response and one of the actions in this step is to prepare a list of all assets and rank them by their level of importance. For example, servers, networks, applications and critical endpoints are all ranked. Each assets' traffic patterns are then monitored to create baselines that can be used for future comparisons. Moreover, in this phase, a communication plan is created, with guidance on who to contact, how they should be contacted and when they should be contacted for a specific incident. In this phase it is also important to set thresholds for determining which security events should be investigated and to actually create the incident response plans for each category of incident. The *Detection and Analysis* phase identifies security incidents, gathers all the details on the incident and performs some incident analysis in order to determine the entry point and the extent of the breach. *Containment* aims to reduce a threat's impact by patching its entry point. *Eradication* aims to remove the threat. *Recovery* aims to return the system to an operational state if it was non-operational or simply return it to business as usual. The *Post-Incident Activity* phase provides the opportunity to learn from experience, in order to better respond to future security events, and analyze the incident-response pairing to identify areas for improvement.

## B. EUROPEAN NETWORK AND INFORMATION SECURITY AGENCY

The European Network and Information Security Agency (ENISA) is a European Union (EU) agency dedicated to preventing and addressing network security and information security problems. [56] proposes a wide-ranging set of measures to boost the level of security in network and information systems. It aims to ensure that EU countries are well-prepared and are ready to handle and respond to cyberattacks through: the designation of competent authorities, the set-up of computer-security incident response teams (CSIRTs) and the adoption of national cybersecurity strategies alongside Incident Response (IR) capabilities in Europe. Moreover, it also establishes EU-level cooperation both at a strategic and technical level and places an obligation on essential-services providers and digital service providers to take appropriate security measures and notify the relevant national authorities to serious incidents.

[56] presented the first piece of EU-wide cybersecurity legislation [57] and provides an analysis of current operational Incident Response (IR) set-ups within the NIS Directive sectors [58]. [58] The EU Cybersecurity Act revamps and strengthens ENISA and establishes an EU-wide cybersecurity certification framework for digital products, services and processes.

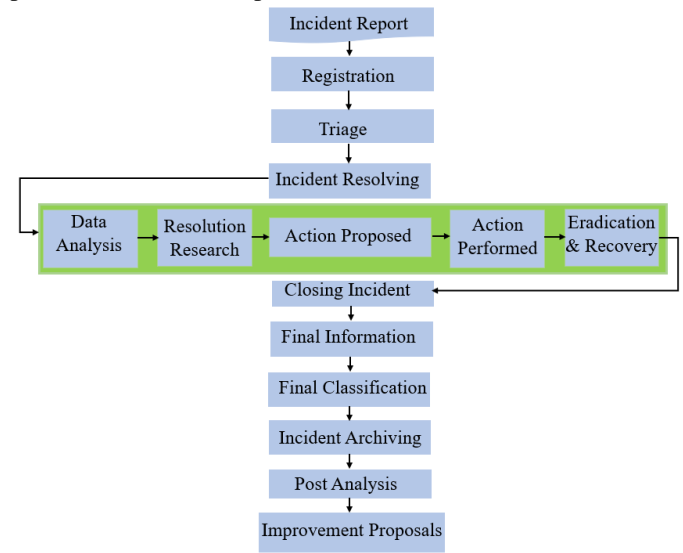

FIGURE 7. *ENISA Incident handling workflow*

A Computer Security Incident Response Team (CSIRT) is a team that receives reports of security breaches, conducts analysis of these reports and responds to the reporters. A CSIRT can be an established group or an ad hoc group of experts. Other widely accepted terms exist for CSIRTs, such as CERT (Computer Emergency Response Team), IRT (Incident Response Team), CIRT (Computer Incident Response Team) or SERT (Security Emergency Response Team). For a comprehensive list of CSIRTs in Europe, ENISA regularly updates an inventory of European

CSIRTs. Furthermore, the Forum of Incident Response and Security Teams (FIRST) and Trusted Introducer (TI) have public links to their global members as well.

[56] also briefly introduces incident response, the main actors in incident response, baseline capabilities these entities should possess in order to effectively combat cyber-attacks and the challenges that hinder efficient incident response. [57] presents the incident handling workflow (as shown in Figure 7), which consists of five phases of incident response (shown in green box): data analysis, resolution research, action proposed, action performed and eradication & recovery.

The first step before *data analysis*, is to inform those who may be the most affected and provide them with information on further proceedings to resolve the incident. The next step is to collect as much relevant data as possible from various sources, such as the incident reporter, monitoring systems and relevant log-files (routers, firewalls, proxy servers, switches, web applications, mail servers, DHCP servers, authentication servers, etc.). In the *resolution research* phase experts exchange ideas, observations and draw conclusions based on the collected data. The *actions proposed* phase prepares a set of concrete and practical tasks for each party. Any action that is proposed should be clear and it should be ensured that the recipient understands exactly what is being proposed. The *action performed* phase is responsible for the proper implementation of all the identified actions in the previous phase. The execution of the actions can be checked by monitoring the network. Finally, the *eradication & recovery* phase helps to restore the service that was affected during the incident.

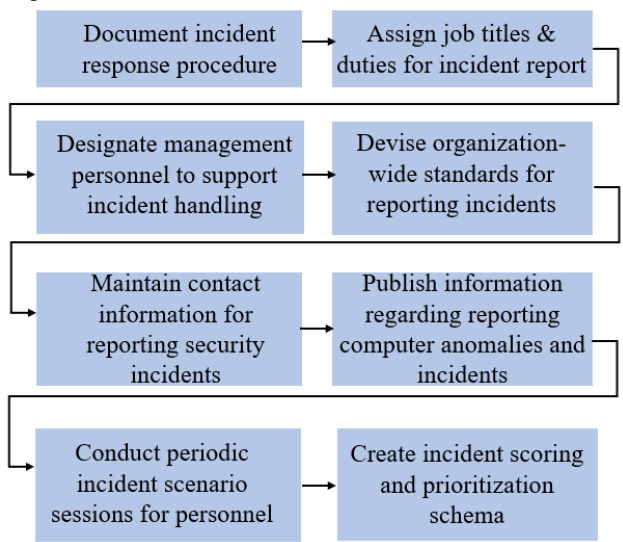

**FIGURE 8.** CIS controls for incident response and management.

ENISA has a key role in setting up and maintaining the European cybersecurity certification framework. This means they have to put in the technical groundwork for specific certification schemes, inform the public on these certification schemes and issue certificates through a dedicated website.

### C. CENTER FOR INTERNET SECURITY

The Center for Internet Security (CIS) publishes the CIS Critical Security Controls in their Controls Assessment Specification document [61]. The controls for incident response and management are presented in 8 steps (shown in Figure 8) in the CIS Control 19 [61]. This includes details on dependencies, inputs, operations, measures and metrics for each step. The first step "*Document incident response procedures*" ensures that there are written incident response plans that define the different roles of personnel in an incident, as well as the different phases in incident handling/management. Step two, "*Assign job titles and duties for incident response*", is based on step one and assigns duties that address tracking and documentation throughout an incident. The third step, "*Designate management personnel to support incident handling*", is also based on step one and identifies management personnel in decision-making roles within incident response, who can help support the personnel responding to an incident. Step four, "*Devise organization-wide standards for reporting incidents*", checks if the Incident Reporting Standards document adequately mentions the time required for system administrators and other members to report anomalous events to the incident handling team, the mechanisms for such reporting and the kind of information that should be included in the notification. Step five, "*Maintain contact information for reporting security incidents*", specifies that the incident response plan should include third-party contact information for reporting security incidents, for example, law enforcement, relevant government departments, vendors and information sharing and analysis partners. Step six, "*Publish information regarding reporting computer anomalies and incidents*", states that the incident response plan should publish incident reporting information to all workforce members as part of the organization's security awareness program. Step seven, "*Conduct periodic incident scenario sessions for personnel*", reviews after action reports and provides a testing ground to check communication channels, decision making and technical competence in the tools and data available. Finally, step eight, "*Create incident scoring and prioritization schema*", determines an incident scoring and prioritization framework based on known or potential impacts, which in turn is used to determine the update frequency of an incident's status and its escalation paths during incident handling.

### D. ISO/IEC 27035: SECURITY INCIDENT RESPONSE

The International Organization for Standardization (ISO) is an international standard-setting body comprised of representatives from various national standards organizations. ISO is the world's largest developer of voluntary international standards and it facilitates world trade by providing common standards for nations. As a result, more than twenty thousand standards have been set, covering everything from manufactured products to technology. ISO has formed two joint committees with the International Electrotechnical Commission (IEC) to develop standards and terminology in the areas of electrical and electronic related technologies.

ISO/IEC 27035 [60] has established information security incident management policies and updates information security and risk management policies. It provides guidelines for building incident management plans by defining technical guidance and other support on establishing an incident response team (CSIRT/CERT). In addition, they also regularly test incident management plans, along with raising security awareness and providing training. Any lessons that are learnt are also recorded for future analysis and guidance. ISO 27035 mentions five phases of incident response as shown in figure 9.

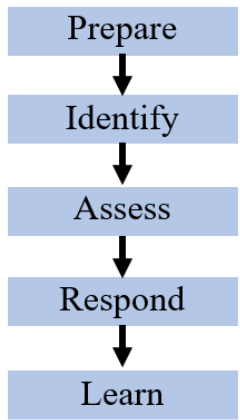

**FIGURE 9.** *Phases of ISO/IEC 27035 Incident Response*

The first is to *prepare* for incidents by formulating incident management policies and establishing a competent team to deal with incidents. Phase 2, *identify*, recognizes and reports information security incidents. The next step is to *assess* incidents and make decisions on how to address them and return to business as usual quickly. The penultimate phase, *respond*, handles incidents by investigating them, containing them and providing a resolution. The final phase is to *learn* from the incident and the responses that were invoked. This includes details on any changes that can improve the overall process.

ISO/IEC 27035 not only outlines, and details approaches to incident management but also explains the benefits of these approaches in helping an organization to fully plan and effectively execute incident management.

As can be seen from the aforementioned guidelines, there are many features that IRSs should consider in order to provide a quick and stable resolution to the detected attack, for example, the "learning step" in each guideline. The last phase of each incident response guideline emphasizes on learning from the process. Most of the IRSs reviewed from the literature do not have learning modules as part of their architecture. These guidelines help in suggesting areas where there are gaps or open challenges for building the next generation of IRSs. The open challenges of IRSs and the author's visions for resolving these issues are described in the next section.

## V. OPEN CHALLENGES AND FUTURE DIRECTION

The basic characteristics of IRSs have shed light on some missing components of IRSs. In this section, some fundamental components/features that IRSs should include to be more effective and efficient are presented. In this article, the effectiveness of an IRS means the response selected and deployed for a detected intrusion produces the intended result (e.g., block the attack and/or mitigate the effect of subsequent attacks), and efficiency of an IRS means achieving the optimal effective response in real time.

### A. INCIDENT AUTHENTICATION

Challenge – The time wasted in generating responses for an attack that was falsely detected by the Intrusion detection system (IDS) reduces the efficiency of the IRS.

IDS [62, 63] provide indicators to potential attacks long before successful attacks are identified. In terms of the accuracy of an IDS, there are four possible states for each activity observed: true positive, true negative, false negative and false positive.

A *true positive* state is when the IDS identifies an activity as an attack and the activity is actually an attack. A *true negative* state is when the IDS identifies an activity as acceptable behavior (non-attack) and the activity is actually acceptable (non-attack). Neither of these states are harmful as the IDS is performing as expected. A *false negative* state is when the IDS identifies an activity as acceptable when the activity is actually an attack. A *false positive* state is when the IDS identifies an activity as an attack, but the activity is an acceptable behavior. A false positive is a false alarm. False positives can cause significant overhead, for example, reducing the effectiveness of security teams by investigating non-malicious events and also IRSs which are handling incidents that are not malicious.

Solution – There should be a component of an IRS that confirms whether an informed incident is actually a true positive. Since if false positive incidents are being handled by an IRS, then IRS efficiency will be reduced since it is trying to handle non-malicious activity. Thus, an *incident authentication* component should be available within an IRS to increase the IRSs competence.

### B. INCIDENT DE-DUPLICATION

Challenge – IRSs informed by various IDSs to the same incident will consider them as different occurrences which reduces the efficiency of the IRS.

IDSs [74,75, 81] can be divided into two types, a

Network based IDS (NIDS) and a Host based IDS (HIDS), depending on the goal of the system. A NIDS monitors network traffic, usually on a mirrored port or in-line, and can be placed in a variety of locations on the network. A HIDS monitors system logs, application logs, and other activities on a host system such as a server or a user's workstation. Most organizations use both types of IDS. They use HIDSs to secure critical host systems and NIDSs to secure their network(s). If multiple IDSs inform an IRS about the same incident then the IRS will consider them as different incidents, so an IRSs efficiency will be negatively affected.

Solution – Thus, an *incident de-duplication* component should be deployed, which will help in reducing the duplicate number of incidents and prevent the system from work overload. If any IDS attempts to create a new incident where there is an open incident that already matches this new alert, then the count value of the incident will be increased by one instead of creating a new incident query and time restrictions on an auto-close policy will be postponed for the specified time of the policy.

### C. INCIDENT PRIORITIZATION

Challenge – IRSs unable to resolve detected incidents that are more urgent or have greater negative impact reduce the effectiveness of the IRS.

An IRS could be informed of many different incidents from many different IDSs at the same time. The decision of which incident should be tackled first is an important issue. An incident's priority level determines the urgency of initiating a response selection procedure and also conveys that response prioritization rank accordingly.

Solution – Thus, *incident prioritization* should be deployed as part of an IRS. Incidents are prioritized according to an incident's severity level.

### D. INCIDENT RESPONSE STORAGE

Challenge – Generating new responses from scratch every time, for every detected incident and not using the lessons learnt from previous responses.

IRSs should use Domain/(Post-)Priori knowledge to select appropriate responses for an incident to save time. This is the knowledge base about the environment in which the incident happens and related facts about the incident. Thus, an IRS should have a database that can store incident and response related information, the incident/response correlation, and even the lessons learnt from previous responses, so as to increase the efficiency of the IRS. The database should be an organized collection of data related to incidents, including information such as, incident information, ranking and corresponding responses with response ranks, positive/negative impact and any extra notes by security analysts for storing the lessons learnt.

Solution – Thus, incident response storage, should be deployed as part of an IRS. Information about incident/response related information or their correlation, along with various factors calculated or informed by domain experts should be stored for future use.

### E. RESPONSE PRIORITIZATION

Challenge – IRSs unable to launch response for incidents that are more urgent or have greater negative impact in time will reduces the effectiveness of the IRS.

After selecting a response for an attack, it should be executed promptly, however the execution of the response depends on other conditions/components and the sequence of the response execution has to be decided. Thus, a response ranking should be considered. Mostly, IRSs execute responses based on a simple First-In-First-Out (FIFO) basis. However, the effectiveness of an IRS can be improved by executing the responses based on their priority. One way of implementing this would be to link a simple logic response priority to the severity of an incident.

Solution – Thus, *response prioritization* should be deployed as part of an IRS.

### F. RESPONSE IMPACT

Challenge – IRSs unable to evaluate the response after deployment reduce the effectiveness and efficiency of IRSs.

The impact of a response should be a fundament feature of an IRS and some may confuse response impact with response evaluation (mentioned in subsection 2.6). The key difference is that response evaluation is used for comparing different responses to select the most appropriate to mitigate an incident. Whereas response impact is a score or feedback once the selected response was deployed. *Response impact* provides the actual result of a deployed response, giving feedback on how successful the response was for handling the incident. Response impact is simply a value calculated to evaluate the effectiveness of the response taken for an attack. It can be calculated based on various factors defined by domain experts. The response impact can help in the selection of responses for future similar incidents and also improve the efficiency of an IRS when stored for future reference.

Solution – Thus, a response impact module that considers both positive and negative impacts of the selected action for detected attacks should be deployed as part of an IRS.

### G. RECOVERY

Challenge – In the aftermath of an attack the compromised system might have deviated from its normal state, which could cause a decrease in the effectiveness of the system.

Normally, Intrusion Prevention Systems (IPS) [71] actively and automatically limit the access to systems for intruders. However, attacks that cannot be blocked by an IPS would be detected by an IDS. Information on system failures and attacks should be passed into an IRS so that it can take an appropriate response to recover from disastrous situations. An IRS should have an appropriate disaster recovery plan too, in order to effectively recover the system and remove any residues of an attack on the system.

Recovery from intrusions is very important, as part of this it is crucial to recover important data that may be lost or manipulated due to intruder activities. *Recovery* systems help with the restoration of a compromised system to a normal state. With the enhancement of high computing power and low-cost hardware and software availability, the cost of human resources outweighs the cost of computing resources, so it is better to have an inbuilt automated intrusion recovery system. As a result, there exists many unaccompanied intrusion recovery systems, but they have an overhead of intrusion exploitation before selecting recovery techniques. Consequently, IRSs containing an inbuilt recovery component will improve the effectiveness of IRSs.

Solution – Thus, a recovery mechanism should be deployed as part of an IRS.

### H. REMOTE MANAGEMENT/HITL

Challenge – Responses automatically generated based on stored information or data could be obsolete or not consider important factors that cause a reduction in the effectiveness of the IRS.

HITL (Human in the loop) means including human feedback into the learning/decision-making process in order to help it improve faster. HITL allows experts to alter the decisions that are made in order to improve the results of the solution that is rectifying a given situation. HITL also allows for the acquisition of knowledge on how a new response may affect a particular situation. HITL is a mix and match approach that may help make an IRS both more efficient and developed. A HITL approach can provide the benefits of both technical knowledge and a method for human intervention in the decision making process. HITL can also be seen as remote management, where experts can change decisions remotely. Specifically, when a data-driven approach is considered, with little data available for decision making, then HITL plays a vital role. In this scenario, HITL helps to maintain human-level precision by incorporating the knowledge of domain experts, which can help to ensure consistency and accuracy. HITL also provides accountability and transparency as decisions can be approved by experts, which in turn increases safety.

Solution – Thus, remote management with human feedback should be deployed as part of an IRS.

### I. HITL/ REMOTE MANAGEMENT

Challenge – Responses automatically generated based on stored information or data could be obsolete or not consider important factors that cause a reduction in the effectiveness of the IRS.

HITL (Human in the loop) means including human feedback into the learning/decision-making process in order to help it improve faster. HITL allows experts to alter the decisions that are made in order to improve the results of the solution that is rectifying a given situation. HITL also allows for the acquisition of knowledge on how a new response may affect a particular situation. HITL is a mix and match approach that may help make an IRS both more efficient and developed. A HITL approach can provide the benefits of both technical knowledge and a method for human intervention in the decision making process. HITL can also be seen as remote management, where experts can change decisions remotely. Specifically, when a data-driven approach is considered, with little data available for decision making, then HITL plays a vital role. In this scenario, HITL helps to maintain human-level precision by incorporating the knowledge of domain experts, which can help to ensure consistency and accuracy. HITL also provides accountability and transparency as decisions can be approved by experts, which in turn increases safety.

Solution – Thus, remote management with human feedback should be deployed as part of an IRS.

IRSs that are also part of an IDRS offer alert verification, however most of the reviewed IRSs trust their respective IDSs for providing verified alerts. Some of the IRSs surveyed provide functionality for recovery and feedback loops, these include ADEPTS [26], Dy-COSIRIS [39] and ACBIRS [49]. Furthermore, IRS RRE [42] and Survivor [54] also have recovery as part of their response. Nevertheless, incident and response prioritization and HITL features could certainly help to increase the efficiency of IRSs.

To build a robust IRS an incident response plan that addresses the aforementioned open challenges, as well as the traditional features of an incident response plan should be drafted. This incident response should address and manage the reaction to an attack or security breach and focus on minimizing damage, reducing disaster recovery time and mitigating breach-related expenses. Hence, a clear, specific and up-to-date incident response plan is a necessity. An incident response plan should address the following questions:

1. Does the incident require a response?
2. Which action should be taken?
3. How and when should an incident response plan be launched?
4. How are the actions going to be performed?
5. How should disaster recovery be performed?

In today's world security breaches (intrusions) cause massive impacts. Therefore, only applying prevention and detection techniques is not sufficient. A proper intrusion response should be prepared for these events to decrease the damage caused by considering the aforementioned solutions to the open challenges.

## VI. CONCLUSION

Intrusion Response Systems are trying to deploy pre-determined response plans, as well as designing new response plans for unknown attacks, based on previous knowledge and the data gathered from new incidents. Even though emerging technological advances like low-cost, high-power computing and fast information processing techniques, such as artificial intelligence and machine learning, have improved intrusion prevention and intrusion detection, there are still significant challenges in using these technologies to the benefit of IRSs.

To understand the complete life cycle of an IRS and to identify the gaps that make current IRSs vulnerable and less efficient, a literature survey has been undertaken by evaluating a number of research papers published during the past decades. A new and complete IRS taxonomy for all phases is presented along with the various techniques that can be used in each phase.

Moreover, a summarized table detailing the specific technology being used for each phase of the surveyed IRSs is presented. Also presented, are the guidelines and standards being proposed by various organizations including NIST, ENISA, CIS and ISO/IEC27035 for incident response management. Finally, future research directions for IRSs based on the results of the literature survey are suggested. For example, improvements in IRS efficiency can be made through incident authentication and de-duplication, improvements in response effectiveness through response impact analysis and recovery and progress in productivity can be achieved through response prioritization and remote management via HITL.

Incidents occur in a wide range of application scenarios from local networks to IoT networks, and there is no one-fit-all solution. From reviewing the literature, IRSs have been identified as multi-phase processes accomplished by various technologies. Robust IRSs require inter-multidisciplinary collaboration of technologies from knowledge-driven to data-driven approaches. As incidents keep evolving, IRSs need to be adaptive, agile and dynamic. The latest work in data analytics and artificial intelligence, e.g., reinforcement learning, are promising candidates for handling the intrusion response in large-scale networks. Moreover, in current IoT environments, IRSs are also subject to the impact of varying government policies between countries. The alignment of IRSs with cybersecurity guidelines is therefore extremely important.

## BIOGRAPHIES

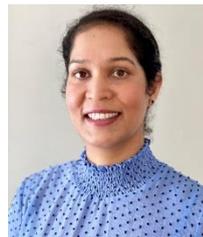

**Pushpinder Kaur Chouhan** (Pushpinder.Kaur.Chouhan@bt.com) is a senior research scientist at BT Plc leading active defense research through deception technology. She received her Ph.D. degree in computer science from ENS-Lyon, France, in 2006. Her current research focuses on future cyber defence that aims to assist security analysts to detect, understand and respond to novel security threats. This involves making use of SDN, AI & ML techniques, and deception technologies to generate insight where it matters most.

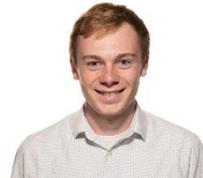

**Alfie Beard** is a research manager for BT leading on autonomous threat response research and has been applying AI to cybersecurity for the past 4 years. He received his MMath in Mathematics in 2018 from the University of Bath. He currently leads the development of Inflame, BT's autonomous threat response and malware simulation tool, which combines deep reinforcement learning for determining the optimal responses to threats and epidemiological modelling for simulating malware propagation.

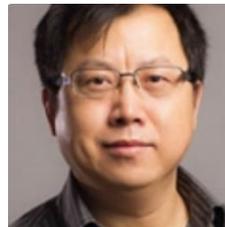

**Liming Chen** received his B.Eng and M.Eng degrees from Beijing Institute of Technology, China, in 1985 and 1988 respectively, and his Ph.D. degree from De Montfort University, UK, in 2003. He is currently Professor of Data Analytics and Research Director for the School of Computing, Ulster University, UK. His research interests include pervasive computing, data analytics, artificial intelligence, user-centred intelligent systems and their applications in smart healthcare and cyber security. He has published over 260 papers in the aforementioned areas. He is an IET Fellow and a Senior Member of IEEE.